\documentclass[floatfix,reprint,prb,showpacs,amsmath,amssymb,superscriptaddress]{revtex4-1}
\usepackage{graphicx}
\usepackage{amssymb}
\usepackage{color}
\usepackage{epstopdf}
\usepackage{bm}

\newcommand{\ket}[1]{\left | \, #1 \right\rangle}
\newcommand{\bra}[1]{\left \langle #1 \, \right |}

\begin{document}
\title{Long-distance spin-spin coupling via floating gates}
\author{Luka Trifunovic}
\affiliation{Department of Physics, University of Basel, Klingelbergstrasse 82,
CH-4056 Basel, Switzerland}
\author{Oliver Dial}
\affiliation{Department of Physics, Harvard University, Cambridge MA, 02138, USA}
\author{Mircea Trif}
\affiliation{Department of Physics, University of Basel, Klingelbergstrasse 82,
CH-4056 Basel, Switzerland}
\affiliation{Department of Physics and Astronomy, University of California, Los
Angeles, California 90095, USA}
\author{James R. Wootton}
\affiliation{Department of Physics, University of Basel, Klingelbergstrasse 82,
CH-4056 Basel, Switzerland}
\author{Rediet Abebe}
\affiliation{Department of Physics, Harvard University, Cambridge MA, 02138, USA}
\author{Amir Yacoby}
\affiliation{Department of Physics, Harvard University, Cambridge MA, 02138, USA}
\author{Daniel Loss}
\affiliation{Department of Physics, University of Basel, Klingelbergstrasse 82,
CH-4056 Basel, Switzerland}
\date{\today}

\begin{abstract} 
The electron spin is a natural two level system that allows a qubit to be
encoded.  When localized in a gate defined quantum dot, the electron spin
provides a promising platform for a future functional quantum computer. The
essential ingredient of any quantum computer is entanglement---between electron
spin qubits---commonly achieved via the exchange interaction. Nevertheless,
there is an immense challenge as to how to scale the system up to include many
qubits.  Here we propose a novel architecture of a large scale quantum computer
based on a realization of long-distance quantum gates between electron spins
localized in quantum dots. The crucial ingredients of such a long-distance
coupling are floating metallic gates that mediate electrostatic coupling over
large distances. We show, both analytically and numerically, that distant
electron spins in an array of quantum dots can be coupled selectively, with
coupling strengths that are larger than the electron spin decay and with
switching times on the order of nanoseconds.
\end{abstract}

\maketitle

\section{Introduction}

Spins of electrons confined to quantum dots provide one of the most promising
platforms for the implementation of a quantum computer in solid state systems.
The last decade has seen steady and remarkable experimental progress in the
quantum control and manipulation of single spins in such nanostructures on very
fast time scales down~\cite{hanson_spins_2007} to 200~ps and with coherence
times of 270~$\mu s$.~\cite{Bluhm_Dephasing_2011}

A large-scale quantum computer must be capable of reaching a system size of
thousands of qubits, in particular to accommodate the overhead for quantum
error correction.~\cite{nielsen:03} This poses serious architectural challenges
for the exchange-based quantum dot scheme,~\cite{loss_quantum_1998}
since---with present day technology---there is hardly enough space to place the
large amount of metallic gates and wires needed to define and to address the
spin qubits. A promising strategy to meet this challenge is to implement
long-range interactions between the qubits which allows the quantum dots to be
moved apart and to create space for the wirings. Based on such a design we
propose a quantum computer architecture that consists of a two-dimensional
lattice of spin-qubits, with nearest neighbor (and beyond) qubit-qubit
interaction. Such an architecture provides the platform to implement the
surface code--the most powerful fault-tolerant quantum error correction scheme
known with an exceptionally large error threshold of
$1.1\%$.~\cite{raussendorf:07,fowler:10}

To achieve such long-range interactions we propose a mechanism for entangling
spin qubits in quantum dots (QDs) based on floating gates and spin-orbit
interaction. The actual system we analyze is composed of two double-QDs which
are not tunnel coupled. The number of electrons in each double-QD can be
controlled efficiently by tuning the potential on the nearby gates. Moreover,
the electrons can be moved from the left to the right dot within each double-QD
by applying strong  bias voltage. Thus, full control over the double-QD is
possible by only electrical means. The double-QDs are separated by a large
distance compared to the their own size so that they can interact only
capacitively. This interaction can be enhanced by using a 'classical'
electromagnetic cavity, i.e., a metallic floating gate suspended over the two
double-QDs, or a shared 2DEG lead between the qubits. The strength of the
coupling mediated by this gate depends on its geometry, as well as on the
position and orientation of the double-QDs underneath the gate. Finally, we
show that spin-qubits based on spins-1/2~\cite{loss_quantum_1998} and on
singlet-triplet states~\cite{levy_universal_2002} can be coupled, and thus
hybrid systems can be formed that combine the advantages of both spin-qubit
types.

\section{Electrostatics of the floating gate}

The Coulomb interaction and spin-orbit interaction (SOI) enable coupling between
spin-qubits of different QD systems in the complete absence of
tunneling.~\cite{flindt_spin-orbit_2006,trif_spin-spin_2007,
stepanenko_quantum_2007,johnson_triplet_2005} However, the Coulomb interaction
is screened at large distances by electrons of the 2DEG and of the metal gates.
Thus, the long-distance coupling between two spin-qubits is not feasible via
direct Coulomb interaction. However, by exploiting long-range electrostatic
forces, it was demonstrated
experimentally~\cite{chan_strongly_2002,kuemmeth_coupling_2008} that QDs can be
coupled and controlled capacitively via floating metallic gates over long
distances. The optimal geometric design of such floating gates should be such
that the induced charge stays as close as possible to the nearest QDs, and does
not spread out uniformly over the entire gate surface. In other words, the
dominant contributions to the total gate-capacitance should come from the
gate-regions that are near the QDs. To achieve a strong qubit-qubit coupling
there is one more requirement: the electric field induced on one QD needs to be
sensitive to the changes of the charge distribution of the other QD. Thus, the
charge gradient, $({\partial q_{ind}}/{\partial \bm r})_{\bm r=0}$, needs to be
large, where $\bm r$ is the position-vector of the point charge with the respect
to the center of the respective QD. To fulfill these requirements we assume the
floating gates consist of two metallic discs of radius $R$ joined by a thin
wire of length $L$. 
\begin{figure}
  \includegraphics[width=8.5cm]{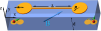}
  \caption{Model system consisting of two identical double-QDs in the $xy$-plane
  and the floating gate between them. The gate consists of two metallic discs of
  radius $R$ connected by a thin wire of length $L$. Each double-QD can
  accommodate one or two electrons, defining the corresponding qubit. Absence of
  tunneling between the separate double-QD is assumed; the purely electrostatic
  interaction between the electrons in the double-QDs leads to an effective
  qubit-qubit coupling. For the spin-$1/2$ qubit the coupling depends
  sensitively on the orientation of the magnetic field $B$. Here $a_0$ is the
  in-plane distance between a QD and the corresponding disc center, while $d$ is
  vertical distance between the QD and the gate.}
  \label{single_gate}
\end{figure}

Let us now investigate the optimal design by modeling the electrostatics of the
floating gates. The electron charge in the QD induces an image charge of
opposite sign on the nearby disc (ellipsoid), see Fig.~\ref{single_gate}. By
virtue of the gate voltage being floating with respect to the ground, the excess
charge is predominantly distributed on the distant metallic ellipsoid, thus
producing an electric field acting on the second QD. In order to carry out the
quantitative analysis of the electrostatic coupling, we  make us of the
expression for an induced charge on the grounded ellipsoidal conductor in the
field of a point charge.~\cite{sten_ellipsoidal_2006} Electrostatic
considerations imply that the coupling (gradient) is enhanced by implementing a
flat-disc design of the gate. Thus, in what follows, we set the disc height to
zero; to reach this regime in practice one only has to ensure that the disc
height be much smaller than its radius. The expression for the induced charge
(in the units of the electron charge) is then given
by~\cite{sten_ellipsoidal_2006}

\begin{eqnarray}
  q_{ind}(\bm r)&=&\frac{2}{\pi}\arcsin(R/\xi_{\bm r}),
  \label{q_ind}
\end{eqnarray}
where $R$ is the
radius of the disc, and $a_0$ is the distance between the QD and the ellipsoid
centers (see Fig. \ref{single_gate}). The ellipsoidal coordinate $\xi_{\bm r}$
is given by
\begin{eqnarray}
  \label{ksir} 2\xi_{\bm r}^2&=&R^2+d^2+|\bm a_0+\bm r|^2\\ &
  &+\sqrt{(R^2+d^2+|\bm a_0+\bm r|^2)^2-4R^2|\bm a_0+\bm r|^2}.  \nonumber
\end{eqnarray}
We emphasize that the induced charge depends only on the coordinate $\xi_{\bm
r}$ of the external charge, as is readily seen from Eq. (\ref{q_ind}). This is
one of the crucial points for the experimental realization of the qubit-qubit
coupling. Thus, positioning the QD below the gate as in  previous
setups~\cite{chan_strongly_2002} is not useful for the qubit-qubit proposed
considered here, since ${\partial q_{ind}}/{\partial \bm r}\approx0$. This fact,
however, can be exploited to turn \textit{on} and \textit{off} the effective
coupling between the qubits. Alternatively, one can use a switch that interrupts
the charge displacement current through the floating gate and thus disables the
build-up of charge gradients at the other disc.

\begin{figure}
  \includegraphics[width=8cm]{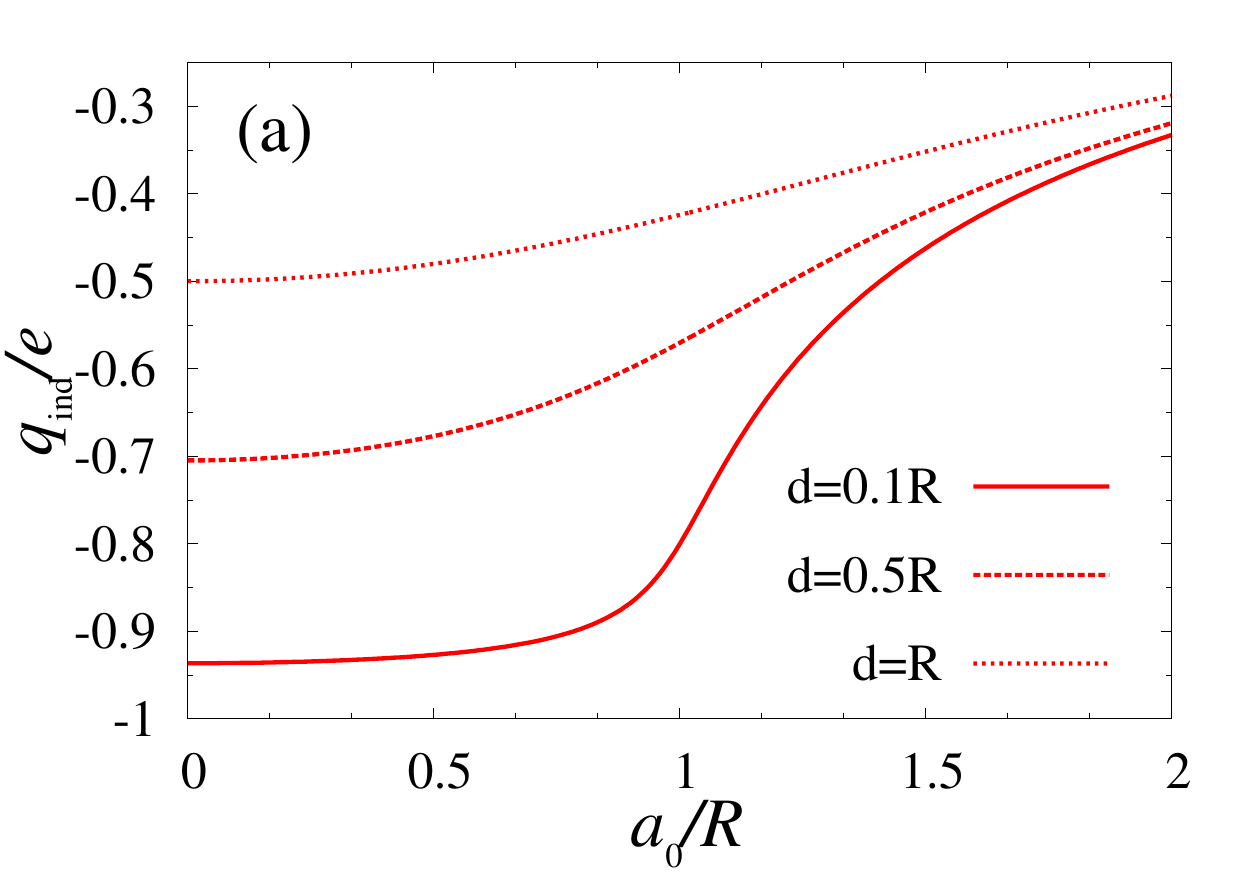}\\
  \includegraphics[width=8cm]{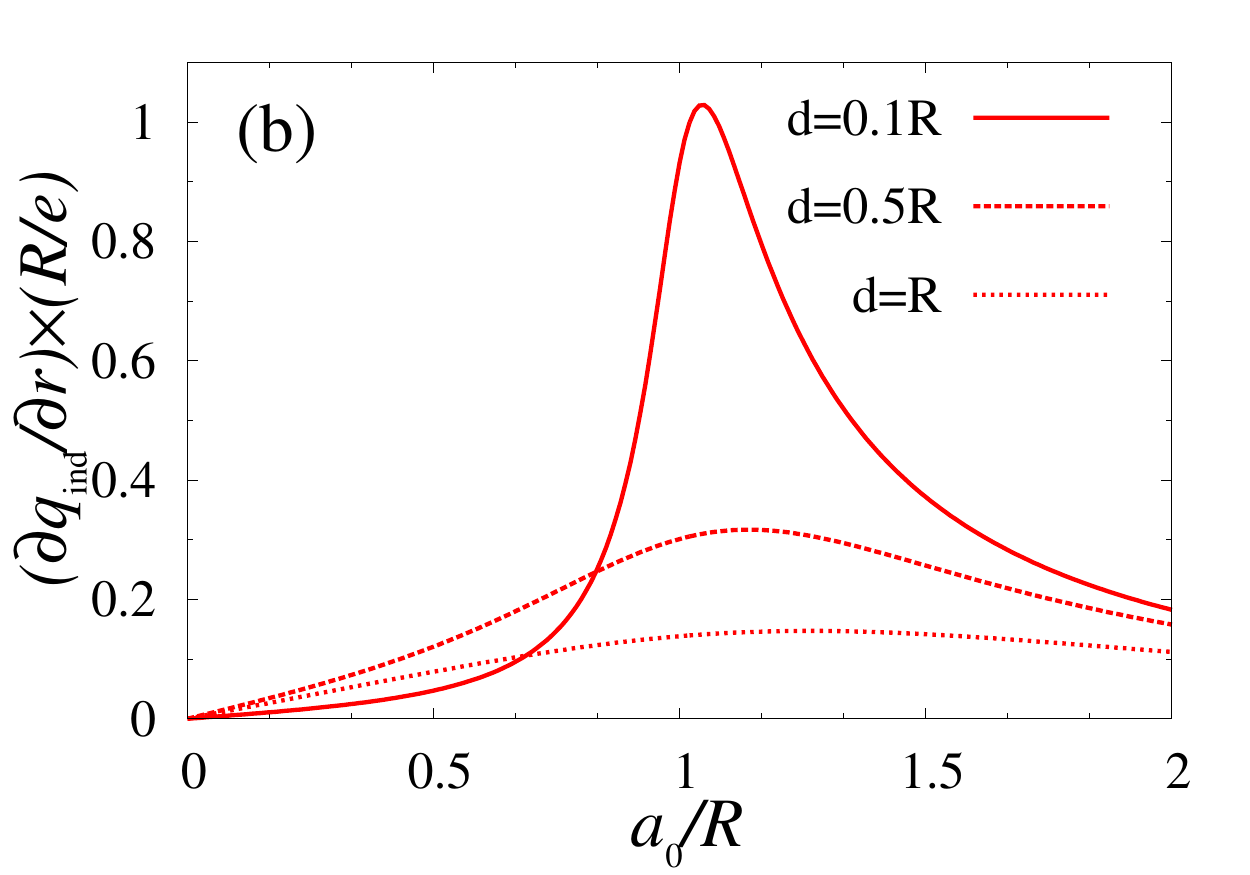}
  \caption{(a) The dependence of the induced charge, $q_{ind}$, and (b) of the
  derivative of the induced charge, $\partial q_{ind}/\partial r$, on $a_0$ at
  $r=0$, i.e. the in-plane distance from the center of the cylindrical gate to
  the center of the QD. We plot these two quantities for several vertical
  distances $d$ between the QD and the gate: $d/R=0.1,\;0.5,\;1$, corresponding
  to the full, dashed and dotted lines, \textit{resp.}.}
  \label{induced_charge}
\end{figure}

Figure \ref{induced_charge} depicts both the induced charged $q_{ind}$, as well
as the charge variation $\partial q_{ind}/\partial r$ as a function of the
horizontal distance $a_0$ between the center of the QD and the center of the
gate. We see that for very small vertical distances $d\ll R$ the variation of
the induced charge peaks at $a_0\approx R$, reaching values as high as unity for
$d=0.1R$, and falls down quickly for $a_0$ larger or smaller than $R$. As
mentioned above, this could be used   as an efficient switching mechanism.
However, as $d$ increases to higher values, comparable to the disc radius $R$,
the charge variation $\partial q_{ind}/\partial r$ flattens out over a wide
range of in-plane distances $a_0$. This means that for larger depths
$d\gtrsim\lambda$ of the quantum dot the  switching mechanism turns out to be
rather inefficient, even though the magnitude of the coupling is only weakly
reduced ($\partial q_{ind}/\partial r\approx0.3$ for $r\approx R$ and $d=0.5R$).
Nevertheless, the gates confining the QDs, as well as the 2DEG itself could lead
to screening of the interaction between the QD and the floating gate, allowing
for an improved switching even in this case. 

Finally, by utilizing the expression for the electrostatic potential of a
charged thin disc~\cite{sten_ellipsoidal_2006} we arrive at the expression for
the electrostatic coupling
\begin{eqnarray}
  V(\bm r_1,\bm r_2)=\frac{\pi\alpha_q}{\kappa}\frac{e^2q_{ind}(\bm
  r_1)q_{ind}(\bm r_2)}{R},
  \label{Vr1r2}
\end{eqnarray}
where $\kappa$ is the dielectric constant, $\alpha_q=\frac{C_d}{C_w+2C_d}$ is
the charge distribution factor of the gate, and $C_d$ and $C_w$ are the
capacitances of the discs and wire, respectively (see Appendix~\ref{A}). We mention that Eq. (\ref{Vr1r2}) is derived in the limit when the
floating gate is immersed in the dielectric,  and it provides a lower bound for
$V(\bm r_1,\bm r_2)$ in the realistic case when the floating gate sits on top of
the dielectric. 

\section{Qubit-qubit coupling}

Next, we consider the coupling between qubits. These can be for either single-
or double-QDs. The two-qubit system with the floating gate is well described by
the Hamiltonian
\begin{equation}
  H  =  V+\sum_{i=1,2}H^{i}_{qubit}\, ,
 \label{H}
\end{equation}
where $V$ describes the electrostatic coupling between the distant charges in
the qubits and is given by Eq. (\ref{Vr1r2}), and $H^{i}_{qubit}$ stands for
either the single-QD or double-QD
Hamiltonian~\cite{burkard_coupled_1999,trif_spin-spin_2007}
\begin{align}
\label{HqubitQD}
  H_{QD}=&H_0+H_Z+H_{SO},\\
  H_{DQD}=&J\:\bm S_1\cdot\bm S_2+H_Z^1+H_Z^2\, .
  \label{HqubitDQD}
\end{align}
Here, $H_0 = p^2_i/2m^*+m^*(\omega_x^2x_i^2+\omega_y^2y_i^2)/2$ is the energy
of an electron in dot $i$ described by a harmonic confinement potential, $m^*$
being the effective mass and $\hbar \omega_{x,y}$ the corresponding
single-particle level spacings. For a single-QD $H_Z=g\mu_B\bm
B\cdot\bm\sigma/2$, stands for the Zeeman coupling, with $\bm\sigma$ the Pauli
matrix for the spin-1/2, and both Rashba and Dresselhaus spin-orbit
interactions
\begin{equation}
  H_{SO}  =  \alpha(p_ x\sigma_y-p_y\sigma_x)
  +\beta(-p_x\sigma_x+p_y\sigma_y),
  \label{Hso}
\end{equation}
where $\alpha$ ($\beta$) is the Rashba (Dresselhauss) spin-orbit interaction
strength. The double-QD is described by an effective Heisenberg
model,~\cite{burkard_coupled_1999} Eq.~(\ref{HqubitDQD}), with $\bm S_i$  being
the spin in the double-QD. In what follows we assume the floating gate to be
aligned along the $x$-axis, see Fig. \ref{single_gate}.

\subsection{Singly occupied double-QDs}

We start by considering two single-QD qubits. Let us first give a physical
description of the qubit-qubit coupling. The purely electrostatic coupling
between the QDs involves only the charge degrees of freedom of the electrons.
Within each QD the spin degree of freedom is then coupled to the one of the
charge via spin-orbit interaction. Hence, we expect the effective spin-spin
coupling to be second order in the SOI and first order in the electrostatic
interaction. In fact, one has also to assume Zeeman splitting to be present on
at least one QD in order to remove the van Vleck
cancellation.~\cite{vleck_paramagnetic_1940,golovach_phonon_2004}

Proceeding to a quantitative description, we assume the spin-orbit strength to
be small compared to the QD confinement energies $\hbar\omega_{x,y}$. Following
Refs.~\onlinecite{golovach_phonon_2004,trif_spin-spin_2007}, we apply a unitary
Schrieffer-Wolff transformation to remove the first order SOI terms. The
resulting Hamiltonian has decoupled spin and orbital degrees of freedom (to
second order in SOI), with the effective qubit-qubit coupling (see Appendix~\ref{A}), with 
\begin{align}
\label{Hssv}
  H_{S-S}&=J_{12}(\bm\sigma_1\cdot\bm\gamma)(\bm\sigma_2\cdot\bm\gamma)\\
  J_{12}=&\frac{m^*\omega_{x,12}^2E_{Z}^2}{2(\omega_x^2-E_{Z}^2)^2},
  \label{Hssve}
\end{align}
where $\bm\gamma=(\beta\cos2\gamma,-\alpha-\beta\sin2\gamma,0)$; $\gamma$ being
the angle between the crystallographic axes of the 2DEG and the $xyz$-coordinate
system defined in Fig.~\ref{single_gate}. Here we assumed for simplicity  that
the magnetic field is perpendicular to the 2DEG substrate,  with $E_{Z}=g\mu_B
B$ the corresponding Zeeman energy (assumed also the same for both dots).
However, neither the orientation nor the possible  difference in the Zeeman
splittings in the two dots affect  the functionality of our scheme (see Appendix~\ref{A} for the most general coupling case). We mention
that  the spin-spin interaction in Eq. (\ref{Hssv}) is of Ising type, which,
together with  single qubit gates forms a set of universal gates (see below).

All information about the floating gate coupling is embodied in the quantity
\begin{equation}
  \omega_{x,12}^2=\pi\alpha_q\alpha_C\left( \frac{\partial q_{ind}}{\partial\tilde x} \right)_{\bm r=0}^2\omega_x^2,
  \label{omega12}
\end{equation}
where $\alpha_C=e^2/(\kappa R\hbar\omega_x)$, and $\tilde x=x/\lambda$
($\lambda$ is the QD size). It is interesting to note that the derived coupling,
Eq.~(\ref{omega12}), is independent of the orbital states of the QDs, and thus,
insensitive to charge fluctuations in the dots. More importantly, the coupling
has only a weak dependence on the wire length $L$---through the capacitance
ratio $\alpha_q$.

Next, we give estimates for the qubit-qubit coupling for GaAs and InAs QDs.
Taking the spin-orbit strength for GaAs semiconductors
$\lambda/\lambda_{SO}\simeq10^{-1}$, and assuming $E_{Z1}\simeq E_{Z2}\equiv
E_Z\simeq0.5\hbar\omega_x$ ($B=2T$ and $\hbar\omega_x\simeq1meV$), we obtain
$H_{s-s}\simeq\alpha_q\alpha_C(\partial q_{ind}/\partial\tilde x)^2_{\bm
r=0}\times10^{-7}eV$. The electrostatic coupling strongly depends (like
$d^{-2}$) on the vertical distance between the gate and the QDs. Typically,
$d\simeq\lambda$, and one obtains using Eq. (\ref{q_ind})  maximal coupling
$H_{s-s}\simeq10^{-11}-10^{-10}eV$ (for $R=1.6\lambda$, $L=10\mu m$, and
$R_w=30nm$ leading to $\alpha_q=0.02$; $a_0=1.9\lambda$). Although, it is
experimentally challenging to decrease $d$ to a value of about $10nm$, the gain
would be a significantly stronger coupling $10^{-9}-10^{-8}eV$ (for
$R=0.17\lambda$ and $a_0=0.2\lambda$). Moreover, if a semiconductor with larger
spin-orbit coupling is used---such as InAs ($\lambda/\lambda_{SO}\simeq1$)---the
coupling is increased by two orders of magnitude compared to GaAs, reaching the
$\mu eV$-regime. Quite remarkably, these values almost reach within the exchange
strengths range, $J_{exc}\sim 10-100 \mu eV$, occurring in typical GaAs double
quantum dots.~\cite{loss_quantum_1998,hanson_spins_2007} Actually, for
realistic devices---as presented in the Sec. \ref{numerics}---the coupling is almost two
orders of magnitude larger then the estimates presented herein. This discrepancy
is not very surprising and it is mainly due to our pessimistic treatment of the
dielectric, and the sensitivity of the electric field gradient to geometry of
the surrounding gates.

\subsection{Hybrid spin-qubits}

A number of different spin-based qubits in quantum dots have been investigated
over the years,~\cite{zak_quantum_2010} each with its own advantages and
challenges. The most prominent ones are spin-1/2 and singlet-triplet spin
qubits. Here, we show that these qubits can be cross-coupled to each other and
thus hybrid spin-qubits can be formed which open up the possibility to take
advantage of the 'best of both worlds'.

We model the hybrid system by a single- and a double-QD qubit, described by 
Eqs.~(\ref{HqubitQD}) and (\ref{HqubitDQD}), respectively. The single-QD and the
floating gate act as an electric field, leading to the change in the splitting
between the logical states of the double-QD spin-qubit, $J\rightarrow J +\tilde
x_e\delta \tilde J $,~\cite{burkard_coupled_1999} with $x_e=\tilde x_e\lambda$
being the $x$-coordinate of the electron in the single-QD and 
\begin{equation}
 \delta \tilde  J = \frac{3}{\sinh(2\tilde l)}
  \frac{\omega_{x,12}^2}{\tilde l\omega_D^2}\epsilon\, .
  \label{efield}
\end{equation}
Here, $\omega_D$ is the confinement energy in the DQD, $\tilde l$ is the
distance between the double-QD minima measured in units of a QD size $\lambda$.
The previous formula is valid for the regime $\epsilon\gtrsim\omega_D$.

In order to decouple spin and orbital degrees of freedom, we again employ a
Schrieffer-Wolff transformation and obtain the hybrid coupling in lowest order
(see Appendix~\ref{B})
\begin{equation}
  H_{hybrid}=\frac{3\mu g\;\delta\tilde J(\bm\gamma\times\bm
  B)\cdot\bm\sigma}{4(\omega_x^2-E_{Z1}^2)\lambda} \tau_z\, .
  \label{HS-s}
\end{equation}
Here, $\tau_z$ is a Pauli matrix acting in the pseudo-spin space spanned by the
logical states of the singlet-triplet qubit. It should be noted that the sign of
this coupling can be manipulated by changing the sign of the detuning voltage
$\epsilon$. As an estimate, we can write
$H_{hybrid}\simeq\left(\frac{\omega_{x,12}}{\omega_x}
\right)^2\frac{E_Z}{\omega_D}\frac{a_B}{\lambda_{SO}}\epsilon$. Assuming the
parameters cited in the previous section for the GaAs-QDs we obtain the estimate
$H_{S-s}\simeq10^{-10}-10^{-9}eV$. Reducing the distance $d$ or using InAs-QDs
we can gain one order of magnitude more in the coupling.

\subsection{Doubly occupied double-QDs}

To complete our discussion about the qubit-qubit couplings, we now consider two
double-QDs coupled via the floating gate. As already noted, owing to the
different charge distributions of the logical states in the double-QD, the SOI
term is not needed for the qubit-qubit coupling.~\cite{stepanenko_quantum_2007}
Certainly, the SOI exists in double-QDs but its effect on the ST splitting can
be neglected. Below only a rough estimate of the
coupling is provided, while the detailed analysis can be found in
Ref.~\onlinecite{stepanenko_quantum_2007}.
%remember ~\textit{(Dimitri's paper)}

We assume both double-QDs to be strongly detuned, thereupon the singlet logic
state is almost entirely localized on the lower potential well of the
double-QD. The electrostatic energy difference between the singlet-singlet and
triplet-triplet system configurations gives the rough estimate of the
qubit-qubit coupling, $H_{S-S}\simeq V(R,R)-V(R+l,R+l)$. Taking the distance
between the double-QD minima $l\simeq R$ and the same GaAs parameters as
before, we finally obtain the estimate $H_{S-S}\simeq10^{-5}-10^{-6}eV$. As can
be seen from Fig. \ref{induced_charge}, reducing $d$ to $10nm$ increases the
coupling five times.

\section{Scalable Architecture}

One central issue in quantum computing is scalability, meaning that the basic
operations such as initialization, readout, single- and two-qubit gates should
not depend on the total number of qubits. In particular, this enables the
implementation of fault-tolerant quantum error correction,~\cite{nielsen:03}
such as surface codes where error thresholds are as large as
$1.1\%$.~\cite{raussendorf:07,fowler:10}

To this end, the architecture of the qubit system becomes of central
importance.~\cite{divincenzo:09} Making use of the electrostatic long-distance
gates presented above, we now discuss two illustrative examples for such
scalable  architectures. 

\subsection{Design with floating metal gates}

\begin{figure}
  \includegraphics[width=8cm]{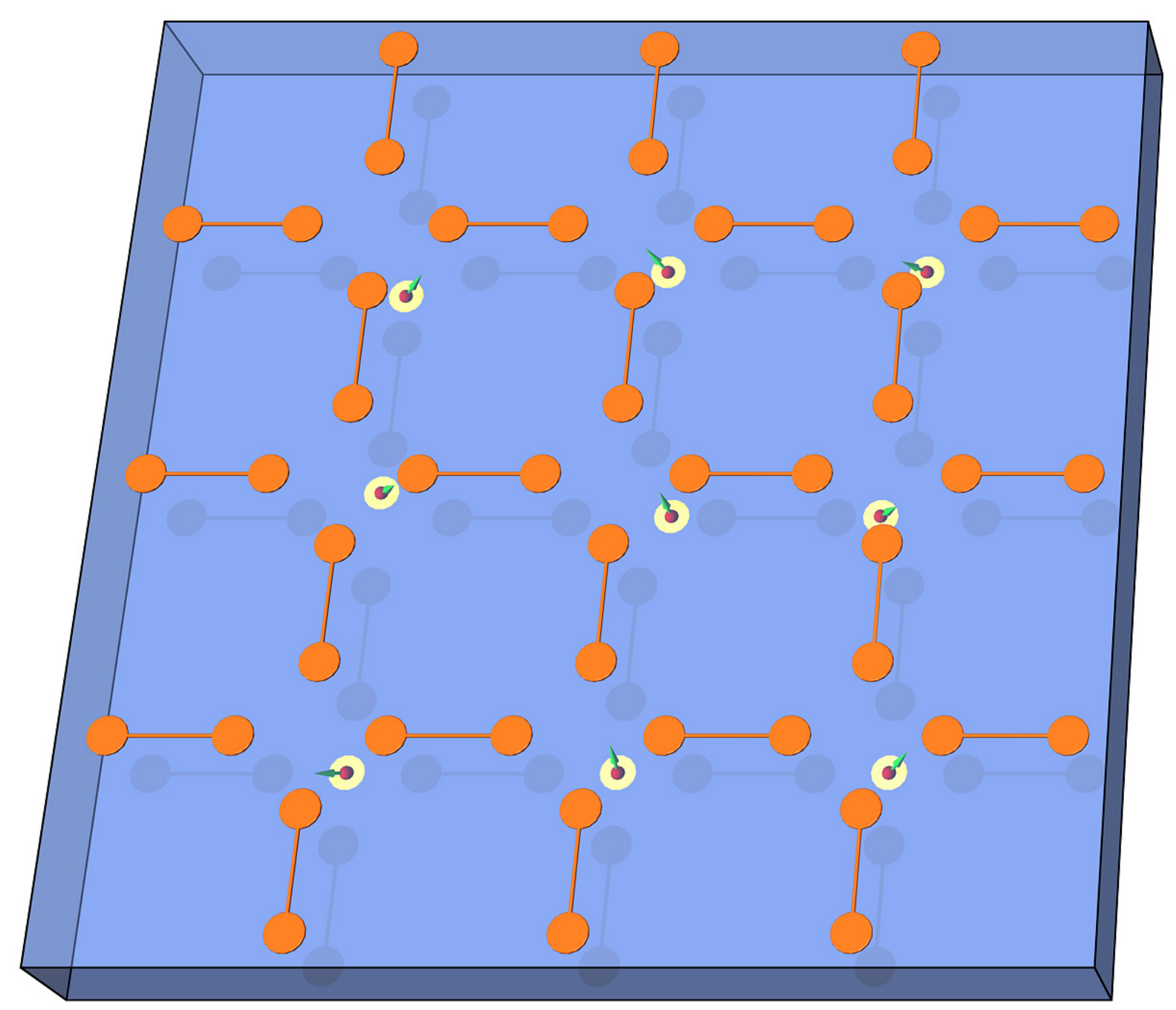}
  \caption{Quantum computer architecture using metallic floating
  gates on top of a 2DEG. The electrostatic long-range coupling is confined to  adjacent
  qubits. Turning on (off) the qubit-qubit interaction is achieved by moving a
  qubit close to (away from) the corresponding metal disc. This architecture allows for parallel switching.}
  \label{qc_metallic}
\end{figure}

In the first design we propose here, the metallic gates above the 2DEG are
utilized for qubit-qubit coupling, while the switching of the coupling is
achieved by moving the QDs (see Fig. \ref{qc_metallic}). Only the coupling
between adjacent QDs is possible in this design. Without this constraint, the
induced charge due to nearby QDs would be spread over the whole system,
resulting in an insufficient qubit-qubit coupling.

The actual virtue of the setup is its experimental feasibility, as suggested by
recent experiments.~\cite{chan_strongly_2002,kuemmeth_coupling_2008} However,
as explained in Sec. II, a minor but crucial difference here is that the
qubit-qubit coupling depends not on the charge itself but rather on its
gradient, in contrast to earlier
designs.~\cite{chan_strongly_2002,kuemmeth_coupling_2008} This requires the
dots to be positioned off the disc-center.

In order to complete our quantum computer design, we have to equip our system
with a fast switch. The discussion in Sec. II is relevant therefore, because
the coupling can be turned \textit{off} (\textit{on}) by moving a QD away
(towards) the corresponding floating gate, see Fig.~\ref{induced_charge}. The
spatial change of the quantum dot induces an electric response in the metallic
floating gate on a time scale roughly given by the elastic mean free time (at
low temperatures). This is the time it takes to reach the new electronic
equilibrium configuration that minimizes the electrostatic energy. Since for a
typical metal this time is on the order of tens of femtoseconds, this response
time poses no limitations, being much faster than the effective switching times
obtained in the previous sections.

\subsection{All-in-2DEG design}

We now consider a setup where are all elements of the qubit-network, including
the floating connector gates, are implemented in the 2DEG itself.  This will
allow us to extend  the above design in an essential way, namely to implement a
switching mechanism inside the connectors themselves which is potentially fast
and  efficient (with a large on/off ratio). There are two attractive features
coming with such a design. First, the qubit-qubit coupling is now controlled by
the connector switch only, while the quantum dots with the spin-qubits  can be
left fixed, thereby reducing the source of gate errors. Second, this design
allows for coupling beyond nearest neighbor qubits, which is beneficial for the
error threshold in fault-tolerant quantum error correction
schemes.~\cite{divincenzo:09}

\begin{figure}
  \includegraphics[width=8cm]{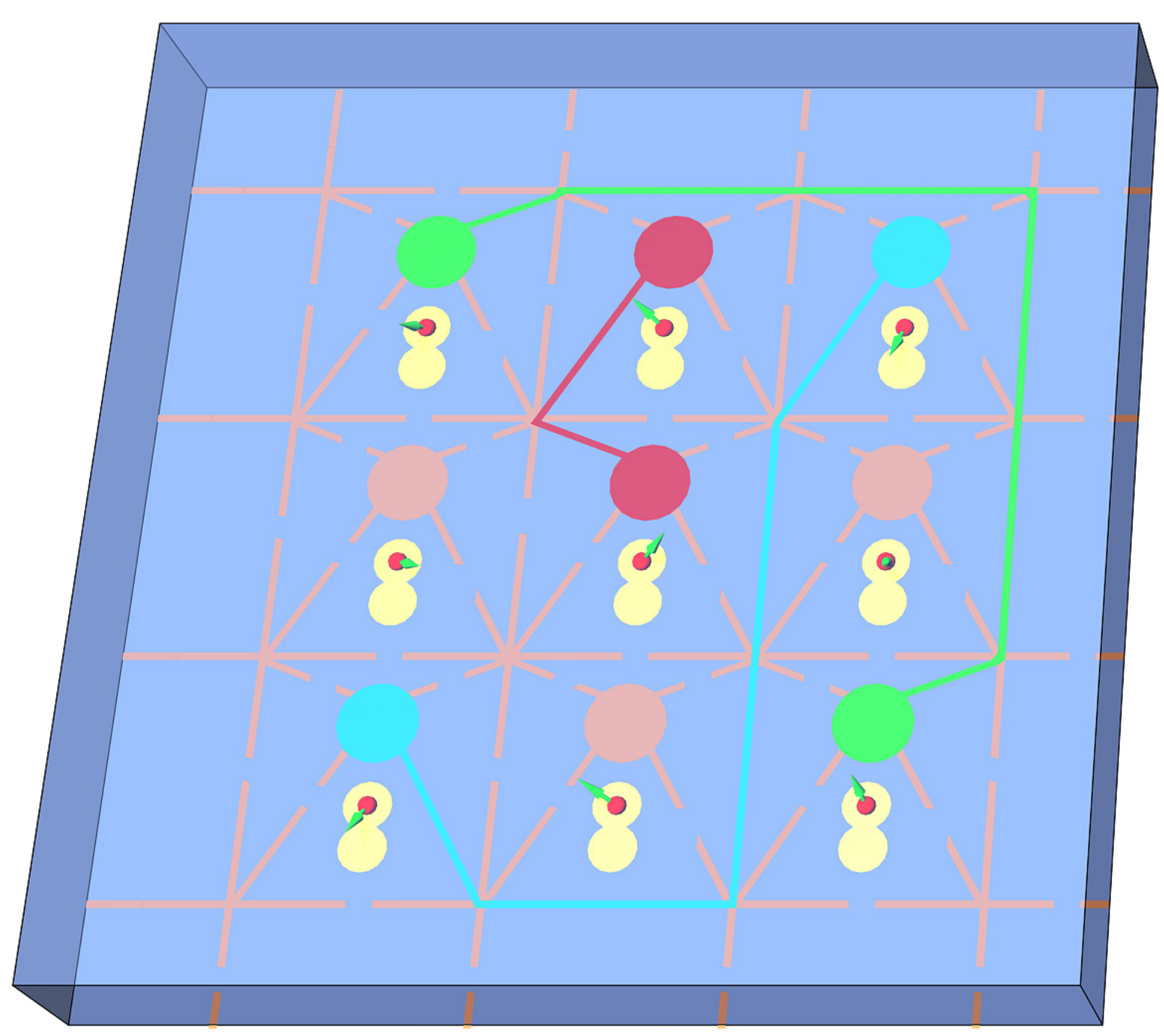}
  \caption{All-in-2DEG design: the qubits and the floating connector gates are
  all implemented within the same 2DEG. The spin-qubits (green arrow) are
  confined to double quantum dots (small yellow double circles) and are at a
  fixed position with maximum coupling strength to the floating gate (big disc)
  (see Fig.~\ref{induced_charge}). The  network consists of quantum channels
  (lines) that  enable the electrostatic coupling between discs (large circles)
  so that two individual qubits at or beyond nearest neighbor sites can be
  selectively coupled to each other. In the figure shown are four pairs of
  particular discs that are connected by quantum channels (full lines), while
  the remaining discs (red) are disconnected from the network (interrupted red
  lines) The discs can be considered as large quantum dots containing many
  electrons. The quantum wires can be efficiently disconnected (interrupted
  lines) by depleting the single-channel with a metallic top gate (not shown).
  This architecture allows for parallel switching.}
  \label{qc_allin2deg}
\end{figure}

The proposed network is shown in  Fig. \ref{qc_allin2deg} where the floating
gates are formed within the 2DEG in form of discs connected by quantum wires.
The discs themselves can be considered as large quantum dots containing many
electrons ($\sim 50-100$) so  that (quantum) fluctuations are negligibly small.
Parts of the network are then connected or disconnected by locally depleting
these wires with the help of a standard quantum point
contact.~\cite{hanson_spins_2007} This suppresses the displacement of charges
very quickly and efficiently.  The electrostatics of such semiconductor gates is
essentially the same as the previously discussed metallic one. Indeed, the
number of electrons in the 2DEG-defined network can be fixed, thus the gate
behaves as floating. Again, the minimal switching time is limited roughly by the
elastic mean free time (at low temperatures), which for a typical GaAs 2DEG is
on the order of tens of picoseconds.

The single spin control required for completing the  universal set of  gates in
our proposal can be implemented in both  setups through  ESR,\cite{KBT+06} or
purely electrically via EDSR,\cite{GBL06,nowack:07,perje:10} which is more
convenient for our electrostatic scheme. The time scales achieved are on the
order of 50 ns, much shorter than the spin relaxation and decoherence
times.\cite{GBL06,nowack:07,perje:10}

\subsection{Design based on 1D nanowire quantum dots}
 
The floating gate architecture efficiency is strongly dependent on the strength
of the SOI experienced by the electrons in the QDs, which have to be large
enough to overcome the spin decoherence rates. InAs nanowires are such such
strong SOI materials, with strengths larger by an order of magnitude than in
GaAs 2DEG.~\cite{fasth:07} Moreover, the electron spins in  QDs created in these
nanowires show long coherence times~\cite{perje:10} and can be controlled
(electrically) on times scales comparable to those found for the electron spin
manipulation in GaAs gate defined QDs.~\cite{perje:10}

\begin{figure}
  \includegraphics[width=7.5cm]{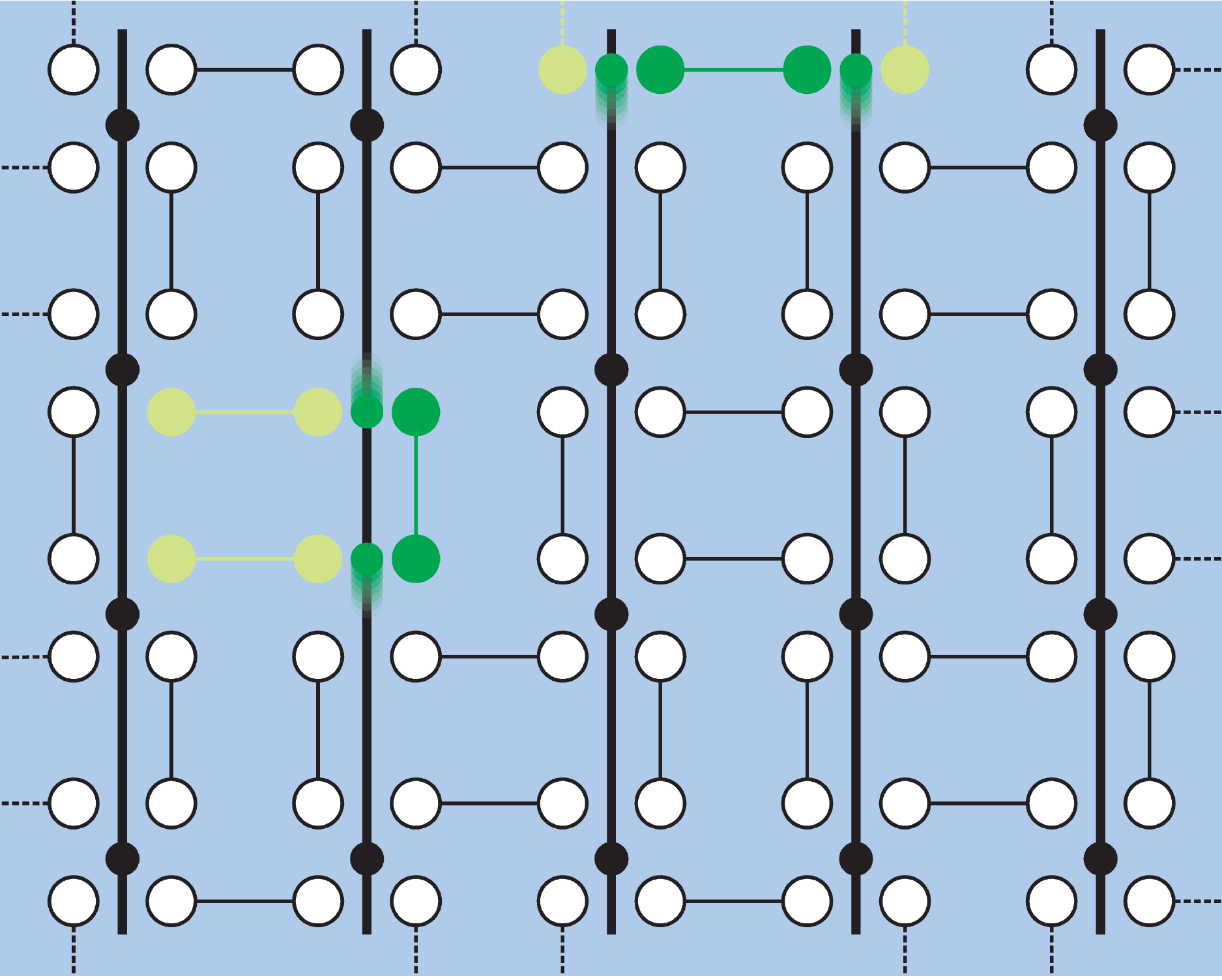}
  \caption{Architecture based on nanowire QDs coupled by metallic gates.  The
  spin qubits are confined to QDs (black dots) on nanowires. The nanowires
  form a parallel array (vertical black lines). The coupling between neighboring
  spin-qubits is enabled by floating metal gates (white) positioned either
  parallel to the wires thus coupling QDs created in the same wire, or
  perpendicular to the wires thus coupling QDs created in adjacent nanowires.
  By using external gates (not shown) to move the dots along the nanowires
  (shaded colors) it is then possible to selectively couple one particular QD to
  only two surrounding gates ('active' gates; green and yellow). The other QD
  partner couple to one of these 'active' gates also (green), thus resulting in
  a selective coupling of the desired nearest neighbor pair.}
  \label{NanowiresSetup}
\end{figure}

In Fig.~\ref{NanowiresSetup} we show a sketch of an architecture based on
nanowires containing single or double QDs. Typical examples for such wires are
InAs~\cite{fasth:07,perje:10} or  Ge/Si~\cite{hu:nna07,kloeffel:11} nanowires,
Carbon nanotubes,~\cite{kuemmeth:08, bulaev:08, churchill:09, klinovaja:prl11}
etc. The default position of a QD is chosen so that the coupling to any of the
surrounding gates is minimal. Neighboring QDs in the same nanowire are coupled
by a vertical metal gate, while  QDs in adjacent nanowires by a horizontal metal
gate. The electron in a given QD can be selectively coupled to only two of the
surrounding gates by moving it (via the gates that confine the electrons) in
regions where the electric field gradient for the induced charge is maximum on
these two 'active' gates,  while negligible for the others two 'passive' gates.
The  other QD partner in the coupling  is moved towards one of the 'active'
gates thus resulting in  a qubit-qubit coupling.  Note that there are in total
three 'active' gates, but only one of them is shared by both QDs, thus allowing
selective coupling of any nearest neighbor pair in the network.
 
The spin coupling  mechanism as well as the 2D geometry are  similar to the
previous 2DEG GaAs QDs designs, showing the great flexibility of the  floating
gate architecture. As before, the spin-qubits can be manipulated purely
electrically, via the same  gates that confine the QDs.~\cite{perje:10} We
mention also that the gate geometry (dog-bone like) shown in
Fig.~\ref{NanowiresSetup} is not optimized to achieve the best switching ratio,
more asymmetric gate geometries possibly leading to better results. 

\subsection{Spin qubit decoherence and relaxation}

Decoherence and relaxation  are  ones of the main obstacles to overcome  in
building a quantum computer.  The main source of qubit decay in typical GaAs
quantum dots comes from nuclear spins and phonons (via spin orbit interaction),
and  has been studied in great detail theoretically and experimentally, see
e.g.~\onlinecite{fischer_dealing_2009}. The longest relaxation and decoherence
times measured are about $T_1\sim 1s$~\cite{amasha_electrical_2008} and $T_2\sim
270\mu s$,~\cite{Bluhm_Dephasing_2011} respectively. Exactly the same qubit
decay mechanisms also apply here, except one new source coming from the Nyquist
noise of the floating metallic gates. However, this problem has been studied in
great detail in Ref.~\onlinecite{marquardt_spin_2005} and no major impact on the
decoherence time was found. Even if Nyquist noise were a problem, it could be
further reduced by using superconducting gates in lieu of normal metal ones.

\section{Implementation of two-qubit gates}

Since the Hamiltonian of Eq. (\ref{Hssv}) is entangling, it can be used to
implement two-qubit gates. Here we consider the CNOT gate, widely used in
schemes for quantum computation.~\cite{raussendorf:07,fowler:10} The
Hamiltonian for two single-QD qubits interacting via the floating gate is the
sum of $H_{S-S}$ and the Zeeman terms. The strength of the latter in comparison
to the former allows us to approximate the Hamiltonian by $H' = J_{12}
|\bm\gamma |^2 (\sigma_x^1 \sigma_x^2 + \sigma_y^1 \sigma_y^2)/2 + E_z
(\sigma_z^1 + \sigma_z^2)/2$, for which qubit-qubit interaction and Zeeman
terms commute. The CNOT gate, $C$, may then be realized with the following
sequences,
\begin{eqnarray} \nonumber
  C &=& \sqrt{\sigma_z^1} \, \sqrt{\sigma_x^2} \, \mathcal{H}^1 e^{i(\sigma_z^1 + \sigma_z^2) E_z t} e^{-i H' t} \\
  && \sigma_x^1 \, e^{i(\sigma_z^1 + \sigma_z^2) E_z t} e^{-i H' t} \, \sigma_x^1 \, \mathcal{H}^1, \\ \nonumber
  C &=& \sqrt{\sigma_z^1} \, \sqrt{\sigma_x^2} \, \mathcal{H}^1 \, \sigma_x^2 \, e^{-i H' t/2} \, \sigma_x^1 \sigma_x^2 \, e^{-i H' t/2} \\
  && \sigma_x^2 \, e^{-i H' t/2} \, \sigma_x^1 \sigma_x^2 \, e^{-i H' t/2} \, \mathcal{H}^1
\end{eqnarray}
where $t = \pi/(4 J_{12} |\bm\gamma |^2)$ and $\mathcal{H}$ denotes the single
qubit Hadamard rotation. These sequences require two and four applications of the floating gate, respectively. More details on their construction can be found in Appendix~\ref{C}. Since $H'$ is only an approximation of the total Hamiltonian, these sequences will yield approximate CNOTs. Their success can be characterized by the fidelity, as defined in Appendix~\ref{C}. For realistic parameters, with the Zeeman terms an order of magnitude stronger than the qubit-qubit coupling, the above sequences yield fidelities of
$99.33\%$ and $99.91\%$ respectively. For two orders of magnitude between the Zeeman terms and qubit-qubit coupling the approximation improves, giving fidelities of $99.993\%$ and $99.998\%$,
respectively. These are all well above the fidelity of $99.17\%$, corresponding
to the threshold for noisy CNOTs in the surface code.~\cite{fowler:10} Hence,
despite the difference in error models, we can be confident that
the gates of our scheme are equally useful for quantum computation.

\section{Numeric Modeling of Realistic Devices}\label{numerics}

\begin{figure}
  \includegraphics[width=8cm]{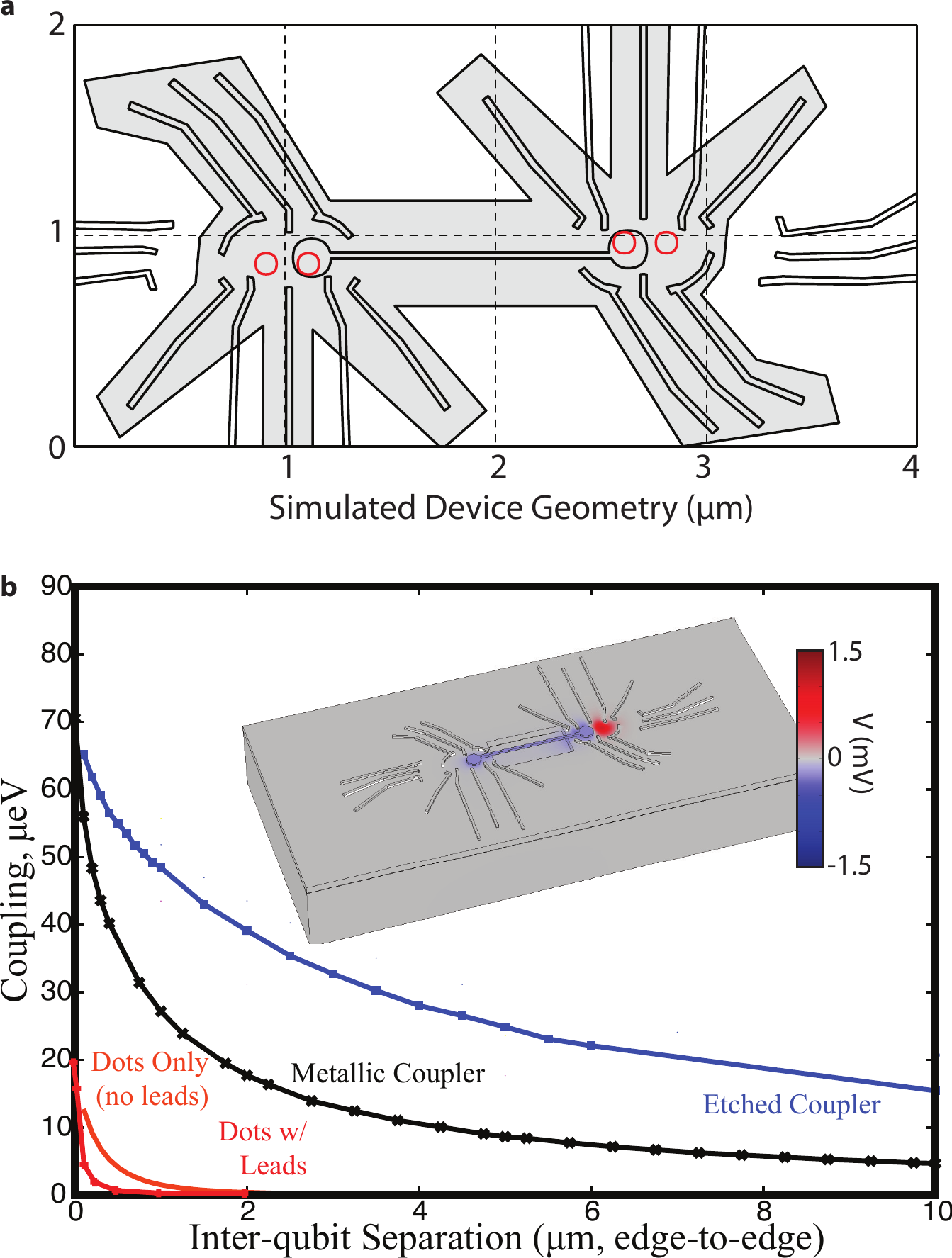}
  \caption{Numeric simulation confirm the efficacy of the design for S-T qubits; 
  addition of a metallic coupler (crosses) increases coupling more than 3-fold for closely
  spaced dots, and greatly extends the range of the coupling. (a) The simulated
  device with a separation of 1 $\mu$m and an etched coupler. 2DEG underneath
  the shaded region is treated as depleted, while red circles show the locations
  of the individual quantum dots within the simulation. (b) Coupling strength as
  a function of separation for the ST qubits in free space (smooth curve),
  qubits including leads and 2DEG but without a coupler (red +), including a
  metallic coupler (black crosses), and additionally etching a trench around the
  coupler to deplete the 2DEG underneath (blue squares). Inset: Electrostatic
  potential (color scale) at the sample surface shows the impact of the coupler
  on a device with a 1 $\mu$m separation.}
  \label{numeric_model}
\end{figure}

In the previous sections, a number of practical concerns related to the
construction of working devices were neglected; most notably, the existence of
the metallic gates used to define the quantum dots themselves and the presence
of undepleted 2DEG outside of the quantum dots. These have finite capacitances
to the coupler, shunting away some of the charge that would otherwise contribute
to the inter-qubit interaction. To confirm that substantial couplings can still
be attained at large distances with these limitations, we have performed numeric
simulations of devices with realistic geometries similar to currently in-use ST
spin qubits. A typical simulated geometry is included in
Fig.~\ref{numeric_model}. The gate and heterostructure design is identical to a
functional device currently being characterized, and the boundaries of the 2DEG
and placement of the electrons within the dot are estimates guided by
experimentally measured parameters. Each quantum dot is modeled as a fixed
charge metallic disc 50 nm in diameter within the 2DEG. While unsophisticated,
this suffices to estimate the practicality of this scheme.

We define the coupling between two ST qubits as the change in detuning in one ST
qubit induced by the transfer of a full electron from one dot to the other dot
in a second ST qubit. For our reference ST qubit design with the two qubits
physically adjacent to each other and no coupler (680 nm center-to-center), we
calculate a coupling of $20 \mu$eV. As the qubits are separated, the coupling
vanishes rapidly as the 2DEG in between the qubits screens the electric field;
it is reduced by an order of magnitude if the dots are separated by an
additional 250 nm. This rapid falloff makes the gate density needed for large
scale integration of these qubits problematic.

Addition of a floating metallic coupler of the type described herein increases
the coupling at zero separation to $70 \mu$eV and allows the qubits to be
separated by more than $6 \mu{}m$ before the coupling drops to the level seen
for two directly adjacent qubits.  We can further improve upon this coupling by
etching the device in the vicinity of the coupler, reducing the shunt
capacitance of the coupler to the grounded 2DEG between the
devices.

\begin{figure}
 \includegraphics[width=8cm]{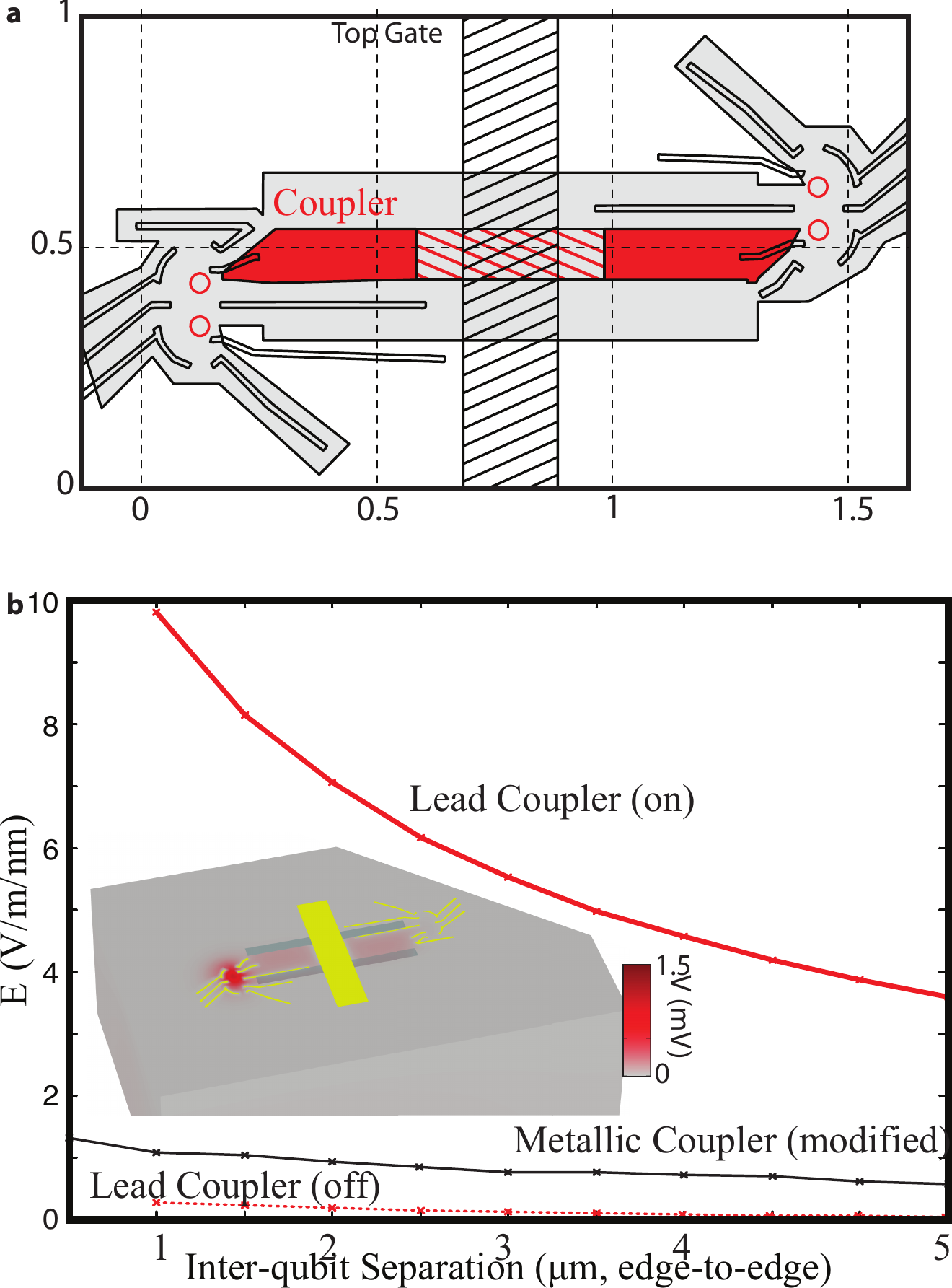}
  \caption{Simulations of single-spin qubits show appreciable coupling strengths,
even over distances of several microns.  While the metallic coupler design of Fig.~\ref{numeric_model} modified to place the quantum dots at the edges of the couplers
  is effective (black crosses in b), an all-in-2DEG design where one of the leads of 
  the qubit acts as a coupler (red region in a) provides dramatically enhanced coupling (solid red lines in b).  The coupler is deactivated by a metallic top gate  (black hatched region in a), modeled by removing the hatched section of the coupler.  Doing so reduces inter-qubit  coupling by over an order of magnitude (dashed red lines in b).  }
  \label{singlespin_model}
\end{figure}

For the case of single spins this metallic coupler is modified to
place the quantum dots at the edges of the coupler rather than under
the discs.  We define the coupling in this case as the electric field in V/m
induced on one qubit in response to 1nm of motion of the electron on the other
qubit.  We continue to find substantial couplings even at
large separations (Fig.~\ref{singlespin_model}).  However, in
this case we find we can further improve couplings by moving to the
all-in-2DEG design where one of the leads of the quantum dot is used
as the coupler (Fig.~\ref{singlespin_model} a).  Using the lead in
this fashion should be harmless; no current is driven into the lead
during qubit manipulations.  The lead (colored region) is modeled as
a metallic strip at the level of the 2DEG.  Due to the close proximity
of the lead to the qubit as well as the sharp electric field gradients
near the point of the lead, we find strongly enhanced coupling for this
lead coupler over the floating metallic coupler for single spin
qubits.  By depleting part of the lead coupler using a metallic top
gate (yellow region), it is possible to selectively turn this coupling
on and off.  The reduction in coupling in the off state is more than
an order of magnitude, and can be further improved by increasing the
size of the depleted region.

\section{Conclusions}

We proposed and analyzed an experimentally feasible setup for implementing
quantum gates in  an array of spin qubits localized in gate-defined quantum dots
based on the interplay of the Coulomb repulsion  between the electrons, SOI and
externally applied magnetic fields. As opposed to the current schemes based on
direct exchange, here there is no need for electron tunneling between the
quantum dots, thus bringing the scheme within experimental reach based only on
current spin-qubit technology.

We showed, both analytically and numerically, that  using either metallic
floating gates in the shape of a dog-bone, or the 2DEG itself acting as a
metallic gate, long-range spin-spin coupling  is achieved, with coupling
strengths exceeding the spin decay rates. Moreover, the coupling can be
selectively switched {\it on} and {\it off} between any pairs of qubits by only
local qubit manipulation, allowing entangling quantum gates such as the CNOT to 
be performed accurately and efficiently. The two-dimensional architecture based on 
the design provides a platform for implementing the powerful surface code.

The electrostatic scheme proposed here is a step forward  towards an efficient
implementation of gates also between hybrid qubits, like ST qubit, hole-spin
qubits, or even superconducting qubits. This  opens up new avenues for a future
working hybrid quantum computer based not on one, but several types of qubits. 

{\em Acknolwedgment.} We thank C. Marcus and D. Stepanenko for helpful
discussions and acknowledge support from the Swiss NF, NCCRs Nanoscience and
QSIT, and DARPA. This research was partially supported by IARPA/MQCO program and 
the U. S. Army Research Office under Contract No. W911NF-11-1-0068. MT acknowledges financial support from NSF under Grant
No.~DMR-0840965.

%%%%%%%%%%%%%%%%%%%%%%%%%%%%%%%
%Appendix
%%%%%%%%%%%%%%%%%%%%%%%%%%%%%%%

\appendix

\section{SPIN-SPIN COUPLING - singly occupied double-dots}\label{A}

In this section we derive explicitly the effective spin-spin coupling. The
spin-orbit interaction (SOI) Hamiltonian $H_{SO}$  is assumed to be small
compared to both the orbital Hamiltonian $H_0+V$ and the Zeeman coupling $H_Z$,
so that we can treat it in perturbation theory. The method of choice for the
perturbation theory is based  on the Schrieffer-Wolff (SW) transformation,
following Refs.~\onlinecite{golovach_phonon_2004,trif_spin-spin_2007}. This
method is very suitable for deriving effective Hamiltonians, as we aim at
herein. We first perform a unitary transformation on the full Hamiltonian, i.e.
$H\rightarrow e^{S}He^{-S}\equiv H_{SW}$, with $S$ an anti-unitary operator so
that we get
\begin{eqnarray}
  H_{SW}&=&H_d+H_{SO}+[S,H_d+H_{SO}]\\
  & &+\frac{1}{2}[S,[S,H_d+H_{SO}]]+\dots,\nonumber
\end{eqnarray}
where $H_d=H_0+V+H_Z$. We look for the transformation $S$ so that this
diagonalizes the full Hamiltonian $H$ in the basis of $H_d$. In leading order in
$H_{SO}$,  we choose $S$ so that $[S,H_0+V+H_Z]=-(1-\mathcal{P})H_{SO}$, with
the projector operator $\mathcal{P}$ satisfying  $\mathcal
PA=\sum_{E_n=E_m}A_{nm}|m\rangle\langle n|$, $\forall A$, i.e. it projects onto
the diagonal part of the Hamiltonian $H_d$. Keeping the lowest order terms in
$\alpha$ and $\beta$  in the SW transformation, we are left with the effective
interaction Hamiltonian $H_{SW}$ that contains the desired spin-spin coupling in
the basis of $H_d$
\begin{eqnarray}
H_{SW}=H_d-\frac{1}{2}\mathcal{P}[S,H_{SO}],
\label{SWHam}
\end{eqnarray} 
where  $S=(1-\mathcal{P})L_d^{-1}H_{SO}$, with $L_d$ being the Liouvillian
superoperator ($L_dA=[H_d,A], \forall A$). 

Next we find the explicit expression for the spin-spin coupling due to the
second-order term in SOI in Eq. (\ref{SWHam}), i.e.
$U\equiv\frac{1}{2}[S,H_{SO}]$. We  make use of the explicit time-dependent
(integral) representation of the Liouvillian $L_d^{-1}=-i\int\limits_0^\infty dt
e^{i(L_d+i\eta)t}$ and arrive at
\begin{equation}
  U=-\frac{i}{2}\int_0^{\infty}dt e^{-\eta t} [H_{SO}(t),H_{SO}],
  \label{U}
\end{equation}
where $H_{SO}(t)=e^{iL_dt}H_n=e^{iH_dt}H_ne^{-iH_dt}$, and $\eta \to 0^+$
ensures the convergence of the time integration. Heisenberg operators, $\bm
\sigma_i(t)$ and $\bm p_i(t)$, are needed in order to calculate $U$. The former
is easy to obtain $\bm\sigma_i(t)=\widehat{\Sigma}_i(t)\bm\sigma_i$, with
$\widehat\Sigma_{i}(t)$ given by
\begin{eqnarray}
  (\widehat{\Sigma}_i)_{mn}(t)&=&\delta_{mn}\bm l_i^2\cos\frac{E_{Zi}t}{2\hbar} +
  2(\bm l_i)_m(\bm l_i)_n\sin^2\frac{E_{Zi}t}{4\hbar}\nonumber\\ &
  &-\varepsilon_{nmk}(\bm l_i)_k \sin\frac{E_{Zn}t}{2\hbar},
  \label{sigmat}
\end{eqnarray}
with $\bm l_i=\bm B_i/B$. The calculation of $\bm p_i(t)$ consists of solving
the system of ordinary differential equations (ODEs)
\begin{eqnarray}
  \frac{d}{dt}\bm p_i(t)&=&-m^*\omega_0^2\bm r_i(t)-\frac{\partial}{\partial\bm
  r_i}V(\bm r_1(t),\bm r_2(t)),\\
  \frac{d}{dt}\bm r_i(t)&=&\bm p_i(t)/m^*.
  \label{p(t)}
\end{eqnarray}

\begin{figure}
  \includegraphics[width=\columnwidth]{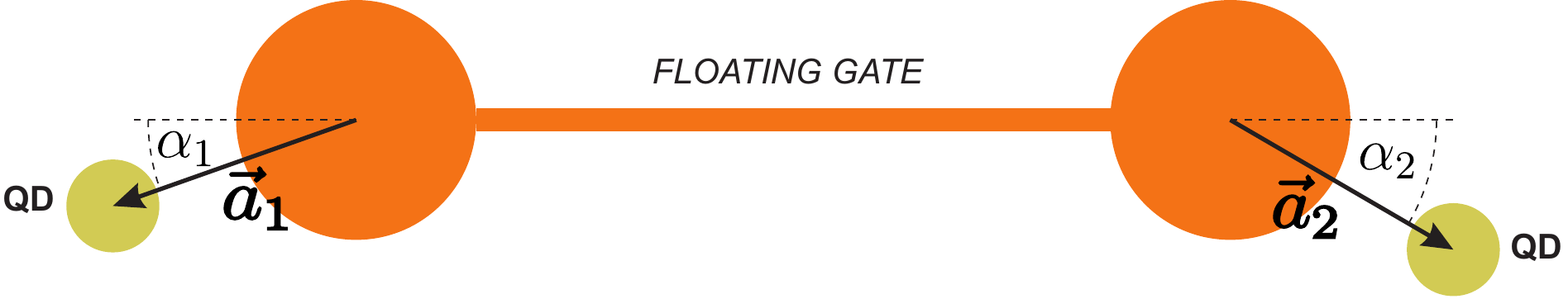}
  \caption{The misalignment angle $\alpha$ of the two QDs (yellow), defined with
  respect to the metallic floating gate (orange).}
  \label{angle_alpha}
\end{figure}

In order to solve this system we expand the electrostatic potential, given in
Eq. (\ref{Vr1r2}), around the minima to second order in $\bm r_i(t)$.
The system of ODEs now reads
\begin{eqnarray}
  \frac{d^2}{dt^2}
  \left(\begin{array}{c}
          \bm p_1(t)\\
          \bm p_2(t)
        \end{array}\right)=-\widehat\Omega
        \left(
        \begin{array}{c}
          \bm p_1(t)\\
          \bm p_2(t)
        \end{array}
        \right), \\
  \widehat\Omega=
    \begin{pmatrix}
      \begin{matrix}
          \omega_x^2 & 0 \\
          0 & \omega_y^2
      \end{matrix}
      & \hat\Omega_{12}\\
      \hat\Omega_{12}^\dagger &
      \begin{matrix}
         \omega_x^2 & 0\\
         0 & \omega_y^2
      \end{matrix}
    \end{pmatrix}.
  \label{pteq}
\end{eqnarray}
In this approximation only terms $O(\bm r_i^2)$ are retained---this is valid for
low lying levels. We ignore the renormalization of the frequencies ($\omega_x$
and $\omega_y$) because it gives higher order (in the Coulomb energy)
contribution to the effective spin-spin coupling. The coupling between
the QDs ($\hat\Omega_{12}$) is given by
\begin{align}
  (\hat\Omega_{12})_{ij}=&\pi\alpha_q\alpha_C\left( \frac{\partial
  q_{ind}}{\partial\tilde{\bm r}_i} \right)\left( \frac{\partial
  q_{ind}}{\partial\tilde{\bm r}_j} \right)\omega_i\omega_j,\label{A_omega12}\\
  ({\partial q_{ind}}/{\partial {\bm r}_i})_{\bm
  r=0}=&\frac{2R\sqrt{\xi_0^2-R^2}{\bm
  a}_i}{\pi\xi_0^2(2\xi_0^2-a_0^2-R^2-d^2)},
  \label{dq_dr}
\end{align}
where $\alpha_q=C_{d}/(C_d+C_w)$, $\alpha_C=e^2/(\kappa R\hbar\omega_x)$, and
$\tilde{\bm r}_i={\bm r}_i/\lambda_i$ ($\lambda_i$ is the QD size along the
$i$-th direction). $\bm a_i$ are the vectors that define the position of the QDs
with respect to the nearby disc center, see Fig. \ref{angle_alpha}. Note that
the expressions for the disc ($C_d$) and wire ($C_w$) capacitances, {\it resp.}
are given by
\begin{eqnarray}
  C_d&=&2 R/\pi,\\
  C_w&=&\frac{L}{2\ln{\left(L/R_w\right)}},
\end{eqnarray} 
where $R$ is the radius of the disk, $R_w$ is the radius of the wire and $L$ is
the length of the wire. 

In order to obtain the solution of Eq. (\ref{pteq}), we note that even a slight
ellipticity ($|\omega_x^2-\omega_y^2|\gg
\max[(\hat\Omega_{12})^2_{xy},(\hat\Omega_{12})^2_{yx}]$) of the QDs causes the
motion in the $x$- and $y$-direction to be decoupled. Having in mind that
$(\hat\Omega_{12})_{yx,xy}^2/\omega_{x,y}^2\sim 10^{-3}-10^{-4}$, we conclude
that such a ellipticity is unavoidable in realistic experimental devices. Thus,
we put off-diagonal elements of the $\hat\Omega_{12}$ matrix to zero and obtain the
solutions
\begin{eqnarray}
  \bm p^i_{1,2}(t)&=&\pm \bm p^i_{xa}\cos(\bm\omega^i_+t)+\bm
   p^i_{xs}\cos(\bm\omega^i_-t)\mp\\
   &\qquad&\mp m^*\bm r^i_a\bm\omega^i_+\sin(\bm\omega^i_+t)-m^*\bm
   r^i_s\bm\omega^i_-\sin(\bm\omega^i_-t),\nonumber
   \label{p(t)sol}
\end{eqnarray}
herein the notation $\bm r_{s,a}=(\bm r_1\pm \bm r_2)/2$, $\bm p_{s,a}=(\bm
p_1\pm \bm p_2)/2$ and $\bm
\omega_\pm=\left(\sqrt{\omega_x^2\pm(\hat\Omega_{12})_{xx}^2},
\sqrt{\omega_y^2\pm(\hat\Omega_{12})_{yy}} \right)$ has been introduced. In the
previous formula, a subscript of a vector denotes the corresponding component of
the vector.

Next, the obtained solutions are inserted into Eq. (\ref{U}). Finally, after
performing the integration over time one obtains the effective spin-spin
coupling for arbitrary orientation of the magnetic field
\begin{eqnarray}
  H_{s-s}&=&\sum_{i=x,y}\frac{m^*\omega_{i,12}^2E_{Z1}^2(\bm l_{1}\times(\bm
  l_{1}\times\bm\gamma_i))\cdot\bm\sigma_1(\bm\sigma_2\cdot\bm\gamma_i)}{4
  (\omega_x^2-E_{Z1}^2)(\omega_x^2-E_{Z2}^2)}\nonumber\\
  & &+1\leftrightarrow2,
  \label{A_Hssv}
\end{eqnarray}
where $\bm\gamma_x=(\beta\cos2\gamma,-\alpha-\beta\cos2\gamma,0)$,
$\bm\gamma_y=(\alpha-\beta\sin2\gamma,-\beta\cos2\gamma,0)$, and $\bm l_i=\bm
B_i/B$. For simplicity of notation, $\bm\gamma_x$ is referred to as $\bm\gamma$ in the main text.

Few remarks should be main herein for the result embodied in the
Eq.~(\ref{A_Hssv}). First of all, from Eq.~(\ref{dq_dr}) we see that
$\hat\Omega_{12}\propto\bm a_1\otimes\bm a_2$, accordingly, the two terms in the
sum of Eq.~(\ref{A_Hssv}) are proportional to $\cos\alpha_1\cos\alpha_2$ and
$\sin\alpha_1\sin\alpha_2$---the angles $\alpha_i$ are being depicted in
Fig.~\ref{angle_alpha}. When only Rashba SOI is present in the material, the
coefficients in front of the two terms are equal and the coupling is
proportional to $\bm a_1\cdot\bm a_2$. This gives yet another efficient
switching mechanism thereby, when the QDs are rotated in such a way that the two
vectors are orthogonal ($\bm a_1\cdot\bm a_2=0$)~\footnote{coupling is zero up
to the small terms $O\left( (\hat\Omega_{12})^2_{xy,yx}/|\omega_x^2-\omega_y^2|
\right)$}.

\section{SPIN-SPIN COUPLING - the hybrid system}\label{B}

We start from the Hamiltonian of the system and then apply
the Schrieffer-Wolff transformation to remove the first order SOI term (present
only in the single QD). The electrostatic potential $V$ is again expanded around
the minimum
\begin{eqnarray}
  \label{Vapprox}
  V(\bm r_e,\bm r_1,\bm r_2)&=&V(\bm r_e,\bm r_1)+V(\bm r_e,\bm
  r_2)\\
  &\approx&m^*\sum_{i=e,1,2}(\delta\omega_x^2 x_i^2+\delta\omega_y^2
  y_i^2)\nonumber\\
  & &+m^*\omega_{x,12}^2 x_e(x_1+x_2),\nonumber
\end{eqnarray}
where $\bm r_e$, $\bm r_1$, and $\bm r_2$ are the coordinates with respect to
the local minima for the electron in the single QD, and the two electrons in the
DQD, respectively. The terms under the sum only renormalize the frequencies, we
do not take them into account, they give only  higher order (in the Coulomb
energy) contributions to the final results. The last term acts as an electric
field on the DQD; as has been shown in the
Ref.~\onlinecite{burkard_coupled_1999}, this leads to a change in the exchange
splitting between the singlet and triplet states in the DQD.
\begin{equation}
  H=H_0+H_Z+H_{SO}+\delta\tilde J \tilde x_e\bm S_1\cdot\bm S_2,
  \label{Hqddqd}
\end{equation}
where $\delta\tilde J$ is given by
\begin{equation}
  \delta\tilde J = \frac{3}{\sinh(2\tilde l^2)}
  \frac{\omega_{x,12}^2}{\tilde l\omega_D^2}\epsilon.
  \label{A_efield}
\end{equation}
$\omega_D$ is the confinement energy in the DQD, $\tilde l$ is the distance
between the DQD minima measured in units of a QD size. We assumed that the
detuning $\epsilon$ is applied to the DQD in order to get the coupling linear in
electrostatic coupling.

The Schrieffer-Wolff transformation is given by $S=(L_0+L_Z+L_H)^{-1}H_{SO}$.
Similarly to the previous section, in order to find the inverse Liouvillian we
have to solve the system of ODEs

\begin{eqnarray}
  \frac{d}{dt} p_{e,x}(t)&=&-m^*\omega_x^2 x_e(t)-m^*\tilde J\bm S_1\cdot \bm S_2,\\
  \frac{d}{dt} p_{e,y}(t)&=&-m^*\omega_y^2 y_e(t),\\
  \frac{d}{dt}\bm r_e(t)&=&\bm p_e(t)/m^*.
\end{eqnarray}
The solution is easily obtained
\begin{eqnarray}
  p^x_e(t)&=&p^x_e\cos(\omega_x t)\\
  & &-m^*\left(x_e\omega_x+\frac{\tilde J}{m^*\omega_x\lambda}\bm S_1\cdot\bm
  S_2 \right)\sin(\omega_x t),\nonumber\\
  p^y_e(t)&=&p^y_e\cos(\omega_y t)-m^*y_e\omega_y\sin(\omega_y t).
  \label{pdqdsol}
\end{eqnarray}
After integration over time, the $S$ transformation is obtained
\begin{eqnarray}
  -iS&=&\sum_{i=x,y}\frac{m^*r_{e,i}\left( \mu^2g^2(\bm
  B\cdot\bm\gamma_i)(\bm B\cdot\bm\sigma)-4\omega_i^2\bm\gamma_i\cdot\bm\sigma
  \right)}{8(\omega_i^2-E_Z^2)}\nonumber\\
  & &+\frac{\mu g(\bm B\times\bm\sigma)\cdot\bm\gamma_i
  p_{e,i}}{4(\omega_i^2-E_Z^2)}\label{S}\\
  & &+\frac{\mu^2g^2(\bm
    B\cdot\bm\gamma_x)(\bm B\cdot\bm\sigma)-4\omega_x^2\bm\gamma_x\cdot\bm\sigma}
    {8\omega_x^2(\omega_x^2-E_Z^2)\lambda}\delta\tilde J\bm S_1\cdot\bm S_2\nonumber,
\end{eqnarray}
The coupling is contained in the $[S,H_Z+\delta\tilde J \tilde x_e\bm
S_1\cdot\bm S_2]$ term 
\begin{equation}
  H_{S-s}=\frac{3\mu g\;\delta\tilde J(\bm\gamma_x\times\bm
  B)\cdot\bm\sigma}{4(\omega_x^2-E_Z^2)\lambda} (\bm S_1\cdot\bm S_2).
  \label{A_HS-s}
\end{equation}
By rewriting the last equation in the pseudo-spin space the generalization for
Eq. (\ref{HS-s}) for arbitrary magnetic field orientation is
obtained.

\section{Implementation of two-qubit gates}\label{C}

Two qubits interacting via the floating gate evolve according to the Hamiltonian
$H = H_{S-S} + E_Z(\sigma_z^1 + \sigma_z^2)$, the sum of the qubit-qubit
coupling and Zeeman term. In general these contributions do not commute, making
it difficult to use the evolution to implement standard entangling gates.
However, when the field is perpendicular to the 2DEG substrate, $H_{S-S}$ takes
the form of Eq. (\ref{Hssv}) which can be decomposed into two terms as follows,
\begin{eqnarray}\nonumber
H_{S-S} &=& {J_{12}}(\Gamma_1-i\Gamma_2 \sigma_z^1) (\sigma_x^1 \sigma_x^2 - \sigma_y^1 \sigma_y^2)/2 \\
&+& {J_{12} |\bf\gamma_x |^2} (\sigma_x^1 \sigma_x^2 + \sigma_y^1 \sigma_y^2)/2.
\end{eqnarray}
Here $\Gamma_1 = ((\bm\gamma_x)_x^2-(\bm\gamma_x)_y^2)$ and $\Gamma_2 =
(\bm\gamma_x)_x(\bm\gamma_x)_y$. The first of these two terms anticommutes with
the Zeeman term, whereas the second commutes. As such, when $E_Z \gg J_{12}
|\bf\gamma_x |^2$, $H_{S-S}$ can be approximated by the second term alone,
\begin{eqnarray}
H_{S-S} &\approx& H_{S-S}' = \frac{J_{12} |\bm\gamma_x |^2}{2} (\sigma_x^1 \sigma_x^2 + \sigma_y^1 \sigma_y^2), \\
H &\approx& H' = H_{S-S}' + E_z (\sigma_z^1 + \sigma_z^2)/2.
\end{eqnarray}
With this approximation, the coupling and Zeeman terms in $H'$ now commute.

We consider the implementation of the gate $\sqrt{\sigma_x \sigma_x} = \exp(-i
\sigma_x^1 \sigma_x^2 \pi/4)$, which is locally equivalent to a CNOT. The
Hamiltonian $H'$ already contains a $\sigma_x^1 \sigma_x^2$ term, so
implementation of the $\sqrt{\sigma_x \sigma_x}$ gate requires only that the
effects of the other terms be removed by appropriate local rotations. Two
possible sequences that can be used to achieve this are,
\begin{eqnarray}\nonumber
   \sqrt{\sigma_x \sigma_x} = &&e^{i(\sigma_z^1 + \sigma_z^2) E_z t} e^{-i H' t}\\ \label{seq1}
&& \sigma_x^1 e^{i(\sigma_z^1 + \sigma_z^2) E_z t} e^{-i H' t} \sigma_x^1 , \\ \nonumber
   \sqrt{\sigma_x \sigma_x} = &&\sigma_x^2 \, e^{-i H' t/2} \, \sigma_x^1 \sigma_x^2 \, e^{-i H' t/2} \\      \label{seq2}
&& \sigma_x^2 \, e^{-i H' t/2} \, \sigma_x^1 \sigma_x^2 \, e^{-i H' t/2},
\end{eqnarray}
where $t = \pi/(4 J_{12} |\bm\gamma |^2)$. The first sequence requires two
applications of the qubit-qubit coupling, whereas the second requires four. The
main difference is that the former removes the effects of the field through the
application of corresponding $z$-rotations after each application of $H'$,
whereas the latter uses $x$-rotations to negate the sign of the field terms and
additional applications of $H'$ to cancel them out. The former is therefore
simpler to implement, however the latter method will also cancel terms not taken
into account in the approximation.

Once the $\sqrt{\sigma_x \sigma_x}$ has been implemented using either of the
above sequences, the CNOT gate, $C$, may be applied using the appropriate local
rotations,
\begin{equation}
C = \sqrt{\sigma_z^1} \, \sqrt{\sigma_x^2} \, \mathcal{H}^1 \,\, \sqrt{\sigma_x \sigma_x} \,\, \mathcal{H}^1.\end{equation}
Here $\mathcal{H}$ denotes the single qubit Hadamard rotation.

Since $H'$ is an approximation of $H$, the above sequences will yield approximate CNOTs, $C'$, when used with the full Hamiltonian. The success of the sequences therefore depends on the fidelity of the gates, $F(C')$. Ideally this would be defined using a minimization over all possible states of two qubits. However, to characterize the fidelity of an imperfect CNOT it is sufficient to consider the following four logical states of two qubits: $\ket{+,0},\ket{+,1},\ket{-,0},$ and $\ket{-,1}$. These are product states which, when acted upon by a perfect CNOT, become the four maximally entangled Bell states $\ket{\Phi^+},\ket{\Psi^+},\ket{\Phi^-},$ and $\ket{\Psi^-}$, respectively. As such, the fidelity of an imperfect CNOT may be defined,
\begin{equation}
F(C') = \min_{i \in \{+,-\}, j \in \{0,1\}} |\bra{i,j} C^{\dagger} C' \ket{i,j}|^2.
\end{equation}
The choice of basis used here ensures that $F(C')$ gives a good characterization
of the properties of $C'$ in comparison to a perfect CNOT, especially for the
required task of generating entanglement.

In a realistic parameter regime it can be expected that $(\bm\gamma_x)_x$ and
$(\bm\gamma_x)_y$ will be of the same order, and the qubit-qubit coupling will
be a few orders of magnitude less than the Zeeman terms. To get a rough idea of
what fidelities can be achieved in such cases using the schemes proposed, we
average over $10^4$ samples for which $(\bm\gamma_x)_y$ is randomly assigned
values between $(\bm\gamma_x)_x/2$ and $3(\bm\gamma_x)_x/2$ according to the
uniform distribution, and $J_{12}(\bm\gamma_x)_x/E_Z=0.1$. This yields values of
$99.33\%$ and $99.91\%$ for the sequences of Eq. (\ref{seq1}) and Eq.
(\ref{seq2}), respectively. For $J_{12}(\bm\gamma_x)_x/E_Z=0.01$ these improve,
becoming $99.993\%$ and $99.998\%$, respectively.

To compare these values to the thresholds found in schemes for quantum
computation, we must first note that imperfect CNOTs in these cases are usually
modelled by the perfect implementation of the gate followed by depolarizing
noise at a certain probability. It is known that such noisy CNOTs can be used
for quantum computation in the surface code if the depolarizing probability is
less than $1.1\%$.~\cite{fowler:10} This corresponds to a fidelity, according to
the definition above, of $99.17\%$. The fidelities that may be achieved in the
schemes proposed here are well above this value and hence, though they do not
correspond to the same noise model, we can expect these gates to be equally
suitable for fault-tolerant quantum computation.

\bibliography{s2s}
\end{document}